\documentclass{aa}  

\usepackage{graphicx}
\usepackage{txfonts}
\usepackage{lipsum}
\usepackage{subcaption}    
\usepackage{lscape}           
\usepackage{placeins}        

\begin{document}

   \title{Turbulent electron transport and asymmetric energy deposition in chromospheric flare ribbons}
   \subtitle{}

   \author{Mehdi Yousefzadeh\inst{1}\email{yousefzadeh@ynao.ac.cn}\corrauth{yousefzadeh@ynao.ac.cn}
           \and Chengcai Shen \inst{1}\corrauth{shenccai@ynao.ac.cn
}\email{shenchengcai@ynao.ac.cn}     
        \and{Alexey Kuznetsov}\inst{2}\email{a_kuzn@iszf.irk.ru}
        \and Marian Karlick\'{y}\inst{3}\email{marian.karlicky@asu.cas.cz}
         \and Yao Chen\inst{4,5}\email{yaochen@sdu.edu.cn}
         \and Mohsen Javaherian \inst{6}\email{moajava@gmail.com}
        \and Mohamed Sedik\inst{1}\email{mohamed.sedik@ynao.ac.cn}
       \and  Abdullah Zafar\inst{1}\email{abdullah@ynao.ac.cn}
         \and Jun Lin\inst{1,7,8}\email{jlin@ynao.ac.cn}
        }

   \institute{Yunnan Observatories, Chinese Academy of Science, Kunming, Yunnan 650216, People's Republic of China
   \and Institute of Solar-Terrestrial Physics, Irkutsk 664033, Russia
   \and Astronomical Institute of the Czech Academy of Sciences, Fri\v{c}ova 298, CZ-25165 Ond\v{r}ejov, Czech Republic
  \and Institute of Frontier and Interdisciplinary Science, Shandong University, Qingdao, Shandong 266237, People’s Republic of China
  \and School of Space Science and Physics, Shandong University, Weihai, Shandong 264209, People’s Republic of China
   \and Research Institute for Astronomy and Astrophysics of Maragha, University of Maragheh, 55136-553 Maragheh, Iran
    \and University of Chinese Academy of Sciences, Beijing 100049, People’s Republic of China
       \and Yunnan Key Laboratory of Solar Physics and Space Science, Kunming 650216, People’s Republic of China
   }

\date{Received: October 07, 2026 - Accepted: November 17, 2026}
 
   \abstract
{Energetic electrons accelerated during solar flares transport energy from the corona to the lower atmosphere, where their precipitation contributes to the heating of flare ribbons. The relative importance of magnetic mirroring and turbulent scattering in controlling this transport remains poorly constrained by observations.}
{We investigate whether the observed energy deposition and asymmetry of flare ribbons can be explained by energetic-electron transport in an asymmetric coronal magnetic loop.}
{We reconstruct the three-dimensional magnetic field of NOAA AR 11283 using a non-linear force-free field extrapolation based on observations and follow few million energetic electrons injected near the loop-top using a guiding-center approximation model. Magnetic mirroring is determined by the extrapolated field geometry, while the effect of turbulence is represented by pitch-angle scattering with characteristic timescales of $\tau_{\mathrm{sc}}= \infty$, $2$, $1$, and $0.5~\mathrm{s}$. The simulated particle population is normalised using the non-thermal electron energy input inferred from observations. We calculate the precipitation rate, cumulative precipitation, ribbon asymmetry, deposited power, and deposited energy flux at the two conjugate foot-points for different scattering frequencies.}
{Magnetic mirroring produces a pronounced precipitation asymmetry, with approximately $80\%$ of precipitating electrons reaching the North ribbon and $20\%$ reaching the South ribbon for no scattering case. Scattering progressively reduces this asymmetry to $64\%/36\%$, $57\%/43\%$, and $52\%/48\%$ for $\tau_{\mathrm{sc}}=2$, $1$, and $0.5~\mathrm{s}$, respectively. The deposited power increases from $\sim4.7\times10^{27}$ to $\sim6.2\times10^{28}~\mathrm{erg\,s^{-1}}$, while the deposited energy flux reaches $\sim8.3\times10^{11}~\mathrm{erg\,s^{-1}\,cm^{-2}}$ as scattering becomes stronger for non-thermal energetic electrons with average energy of around 20 $keV$.}
{The relatively weak turbulent case simultaneously reproduces the observed non-thermal energy deposition, energy flux, and the North-South precipitation partition, consistent with the observed hot-ribbon emission. This agreement provides encouraging evidence that turbulent pitch-angle scattering is crucial in regulating flare-ribbon asymmetry and energy deposition. These results provide an observationally constrained transport framework for subsequent modelling of the thermal response, emission measure, and EUV emission of flare ribbons.}


   \maketitle
   \nolinenumbers

\section{Introduction}
\label{sec:introduction}

Solar flares are among the most energetic transient phenomena in the solar atmosphere. They are powered by the rapid conversion of magnetic energy stored in the corona, most likely through magnetic reconnection, into plasma heating, bulk flows, particle acceleration, and electromagnetic radiation (\cite{Shibata2011,Fletcher2011,Emslie2012}). Although the primary energy release occurs in the corona, a substantial fraction of the released energy is transported toward the denser lower atmosphere. Understanding how this energy is transferred from the coronal magnetic field to the chromosphere is therefore central to determining how flares convert magnetic energy into observable radiation and thermal plasma.

Energetic electrons are thought to be an important carrier of this flare energy. Following acceleration in or near the reconnecting coronal magnetic field, electrons can propagate rapidly along magnetic field lines and transport energy toward the chromosphere. In the tenuous corona, their collisional energy losses can remain relatively small over the characteristic length of a coronal loop, whereas collisions become increasingly important as the electrons encounter the denser lower atmosphere. The resulting energy deposition can produce hard X-ray bremsstrahlung and heat the ambient chromospheric plasma (\cite{Brown1971, Emslie1978, Holman2011, Kontar2011}). The enhanced emission produced by this heating is observed as flare ribbons at the two conjugate foot-points of reconnecting coronal magnetic flux tubes. Consequently, flare ribbons provide an important observational link between the transport of energetic particles in the corona and the response of the lower atmosphere.

Hard X-ray observations provide a direct diagnostic of the accelerated electron population and can be used to estimate the energy carried by non-thermal electrons. Within the collisional thick-target framework, the observed hard X-ray spectrum can be related to the injected electron distribution and its associated energy deposition rate  (\cite{Brown1971,Holman2011,Kontar2011}). However, the inferred electron energy content is sensitive to the low-energy part of the electron spectrum, which is often poorly constrained by hard X-ray observations. Therefore, estimates based on the observed non-thermal component can provide important constraints on the energy input to flare ribbons, but they do not by themselves determine how the electrons are transported from the acceleration region to the two conjugate foot-points.

The spatial distribution of this energy deposition is particularly important in two-ribbon flares. Although the two foot-points are connected by the same coronal magnetic structure, their magnetic field strengths and convergence can differ substantially. An energetic electron moving along a converging magnetic field therefore experiences magnetic mirroring, with the fraction of particles entering the loss cone depending on the magnetic field ratio between the loop-top and each foot-point (\cite{Yousefzadeh2025,Yousefzadeh2026}). Even for an initially symmetric electron population, differences in the two magnetic mirror ratios can consequently produce asymmetric precipitation and energy deposition between the conjugate ribbons. This provides a natural mechanism through which the three-dimensional magnetic configuration can influence the observed distribution of flare heating.
Energetic electron transport during turbulent solar flares involves more than just magnetic mirroring. Pitch-angle scattering also plays a crucial role by redirecting electrons into the loss cone, enabling them to reach the chromosphere despite magnetic trapping. 
A transport model that includes magnetic convergence and pitch-angle scattering is therefore needed to determine how the magnetic field and turbulent scattering jointly control the electron precipitation and thus energy deposited at each ribbon.

Observations of the 2011 September 6 X2.1 flare provide a particularly useful framework for investigating this problem. \cite{Krucker2021} analysed this flare using desaturated extreme-ultraviolet (EUV) observations from the Solar Dynamics Observatory/Atmospheric Imaging Assembly (SDO/AIA), together with hard X-ray observations from the Reuven Ramaty High Energy Solar Spectroscopic Imager (RHESSI; \cite{Lin2002}), to quantify the thermal and non-thermal properties of the flare ribbons. Their differential emission measure analysis showed that the ribbon emission measure increases toward temperatures of approximately $9\times10^{6}~\mathrm{K}$ before declining at higher temperatures, whereas the associated coronal flare loop reaches temperatures of approximately $27\times10^{6}~\mathrm{K}$. The authors further estimated a non-thermal electron energy deposition rate of at least $\sim6.2\times10^{28}~\mathrm{erg\,s^{-1}}$, indicating that accelerated electrons provide sufficient energy input powering the observed flare ribbon heating.

The observational constraints therefore motivate a direct investigation of whether the measured ribbon heating and asymmetry can be reproduced by electron transport through a realistic three-dimensional coronal magnetic field. In particular, it is necessary to determine whether the magnetic geometry alone can account for the observed partition of energetic electrons between the two ribbons, or whether additional turbulent transport is required. A numerical treatment that follows the individual trajectories of energetic electrons in the reconstructed magnetic field provides a means of addressing this question while retaining the spatial information that is lost in spatially averaged transport models.

The principal objective of this work is therefore to establish a direct quantitative connection between  turbulent electron transport, magnetic-field structure, and the energy deposited in flare ribbons. We specifically investigate whether the observed ribbon asymmetry and non-thermal energy deposition can be reproduced by a combination of realistic magnetic-field geometry and turbulent pitch-angle scattering. 
By connecting particle transport in a three-dimensional magnetic field with independently derived flare-ribbon properties, this study provides a framework for interpreting ribbon heating in terms of the underlying transport of energetic electrons.

\section{Observational constraints}
\label{sec:observational_constraints}

This study builds on observational constraints from the GOES X2.1 flare on September 6, 2011, as analysed by \citet{Krucker2021}. This event was selected because its two flare ribbons were observed simultaneously in the EUV and hard X-rays. The EUV ribbon emission was reconstructed from saturated SDO/AIA images using a desaturation technique, while RHESSI observations were used to characterise the hot coronal source and the non-thermal electron population.

\subsection{RHESSI non-thermal electron properties}
\label{subsec:rhessi_non-thermal}

RHESSI observations showed a strong non-thermal hard X-ray component above
approximately \(22~\mathrm{keV}\). The photon spectrum was described by a hard
power law with photon indices reaching values of approximately \(2.9\). Under the
cold thick-target interpretation, the inferred energy deposition rate by
flare-accelerated electrons with energies above \(20~\mathrm{keV}\) was

\begin{equation}
P_{\mathrm{obs}> 20\,\mathrm{keV}}
\simeq 6.2 \times 10^{28}~\mathrm{erg~s^{-1}}.
\label{eq:rhessi_power}
\end{equation}

The value in Equation~\eqref{eq:rhessi_power} is a lower limit, representing the total electron energy deposition for the observed flare. RHESSI could not reliably constrain the non-thermal electron spectrum below \(20~\mathrm{keV}\), and therefore the total electron power may be larger if a substantial population of lower-energy electrons was present. 
The characteristic electron stopping energy was estimated as 

\begin{equation}
E_{\mathrm{stop}} \approx \left(\frac{N_{\mathrm{EUV}}}
 {1.5\times10^{17}~\mathrm{cm^{-2}}} \right)^{1/2} \mathrm{keV},
\label{eq:stopping_energy}
\end{equation}

following \citet{Emslie1978}. Here, $N_{\mathrm{EUV}}$ denotes the electron column density through the hot EUV-emitting ribbon. Following \citet{Krucker2021}, we adopt the column-density estimate derived from the EUV emission-measure/DEM analysis together with the assumed geometry of the emitting region, rather than from the length of the coronal half-loop. The uncertainty in the assumed emitting-region geometry contributes to the uncertainty in $N_{\mathrm{EUV}}$, resulting in the $E_{\mathrm{stop}} \approx 21\mbox{--}24~\mathrm{keV}$ range of characteristic stopping energies.

In the simplest collision-dominated transport picture, electrons below this energy can deposit most of their energy within the hot EUV source, whereas higher-energy electrons can pass through it and deposit energy at greater depths. Pitch-angle scattering and magnetic mirroring can modify this result by increasing the effective path length and changing the fraction of electrons that reaches each ribbon.

The total flare-integrated power in Equation~\eqref{eq:rhessi_power} was not independently separated between the northern and southern ribbons. It therefore provides a constraint on the total energy deposited by the simulated electron population,

\begin{equation}
P_{\mathrm{sim}}^{N}(t)
+
P_{\mathrm{sim}}^{S}(t)
\simeq
P_{\mathrm{obs}>20\,\mathrm{keV}},
\label{eq:intro_total_power}
\end{equation}

where \(P_{\mathrm{sim}}^{N}\) and \(P_{\mathrm{sim}}^{S}\) are the simulated deposition rates in the northern and southern ribbons, respectively. This work examines how magnetic connectivity and electron transport determine the distribution of total power between two ribbons.

\subsection{Flare ribbon geometry and energy deposition}
\label{subsec:ribbon_geometry}

The flare displayed a standard two-ribbon morphology, providing useful geometric constraints for particle-transport calculations. Following \cite{Krucker2021}, we selected the flare-loop arcade as a semicircular magnetic structure, with an approximate foot-point separation of \(25^{\prime\prime}\), resulting in a characteristic full arc length of about \(40^{\prime\prime}\).
RHESSI observed non-thermal emission (\(30\)--\(80~\mathrm{keV}\)) originated from two elongated sources, measuring about $\sim$ $7''$ long and $\sim$ $1''$ wide, spatially aligned with the EUV ribbons.
Based on this geometry, we adopt the combined ribbon area $A_{\mathrm{ribbon}} \simeq 7.4 \times 10^{16}~\mathrm{cm^{2}}$.
Since the radial extent of the emitting plasma could not be directly measured, this area is used to estimate the characteristic thermal energy and convert the total RHESSI-derived non-thermal power into an area-averaged energy flux:

\begin{equation}
F_{\mathrm{obs}} = \frac{P_{\mathrm{obs}>20\,\mathrm{keV}}} {A_{\mathrm{ribbon}}} \approx 8.4\times10^{11} ~\mathrm{erg~cm^{-2}~s^{-1}}.
\label{eq:energy_flux}
\end{equation}

The area-averaged energy flux is constrained by the total non-thermal power, derived from flare-integrated RHESSI spectroscopy, and the AIA-derived ribbon geometry.
Ribbon-specific power estimates would require dedicated RHESSI imaging-spectroscopy analysis, which represents a valuable future extension beyond the scope of this study.

For a numerical model, the energy flux at each ribbon should instead be computed from the local simulated power and the corresponding ribbon area,
\begin{equation}
F_{j}(t) = \frac{P_{j}(t)}{A_{j}},
\qquad
j\in\{N,S\}.
\label{eq:intro_ribbon_flux}
\end{equation}

The observational study does not provide an independent north--south partition of the RHESSI-derived non-thermal power. Therefore, the simulated ribbon asymmetry should be interpreted as a prediction of the transport model rather than as a directly measured observational quantity. 
The hot emission measures derived by \citet{Krucker2021} are \(\mathrm{EM_N}=2.5\times10^{48}~\mathrm{cm^{-3}}\) and \(\mathrm{EM_S}=1.9\times10^{48}~\mathrm{cm^{-3}}\) for the northern and southern ribbon sources, respectively. This ratio corresponds to approximately 57\% and 43\% of the combined hot emission measure in the northern and southern ribbons. Although this ratio is a thermal constraint rather than a direct measurement of the precipitating electron partition, it provides a useful reference for evaluating the modelled north--south precipitation asymmetry.

In the present study, these constraints are used to evaluate whether the present model can simultaneously reproduce the total non-thermal deposited power, the characteristic ribbon energy flux, and the relative precipitation rate between the two magnetic foot-points.

\section{Numerical model}
The numerical framework developed in this study combines a three-dimensional reconstruction of the coronal magnetic field with guiding-center approximation (GCA) simulations to investigate the transport and precipitation of energetic electrons during solar flares.

\subsection{Magnetic field extrapolation}
 The direct measurements of the coronal magnetic field remain challenging, so the field configuration is reconstructed using the non-linear force-free field (NLFFF) extrapolation method based on photospheric observations.
 The input magnetic boundary condition is obtained from the vector magnetograms provided by the Helioseismic and Magnetic Imager (HMI; \cite{Schou2012}) onboard the SDO, which provides full-vector magnetic field measurements with a cadence of approximately $12$ minutes.

\begin{figure*}[ht!]
    \resizebox{\hsize}{!}
             {\includegraphics {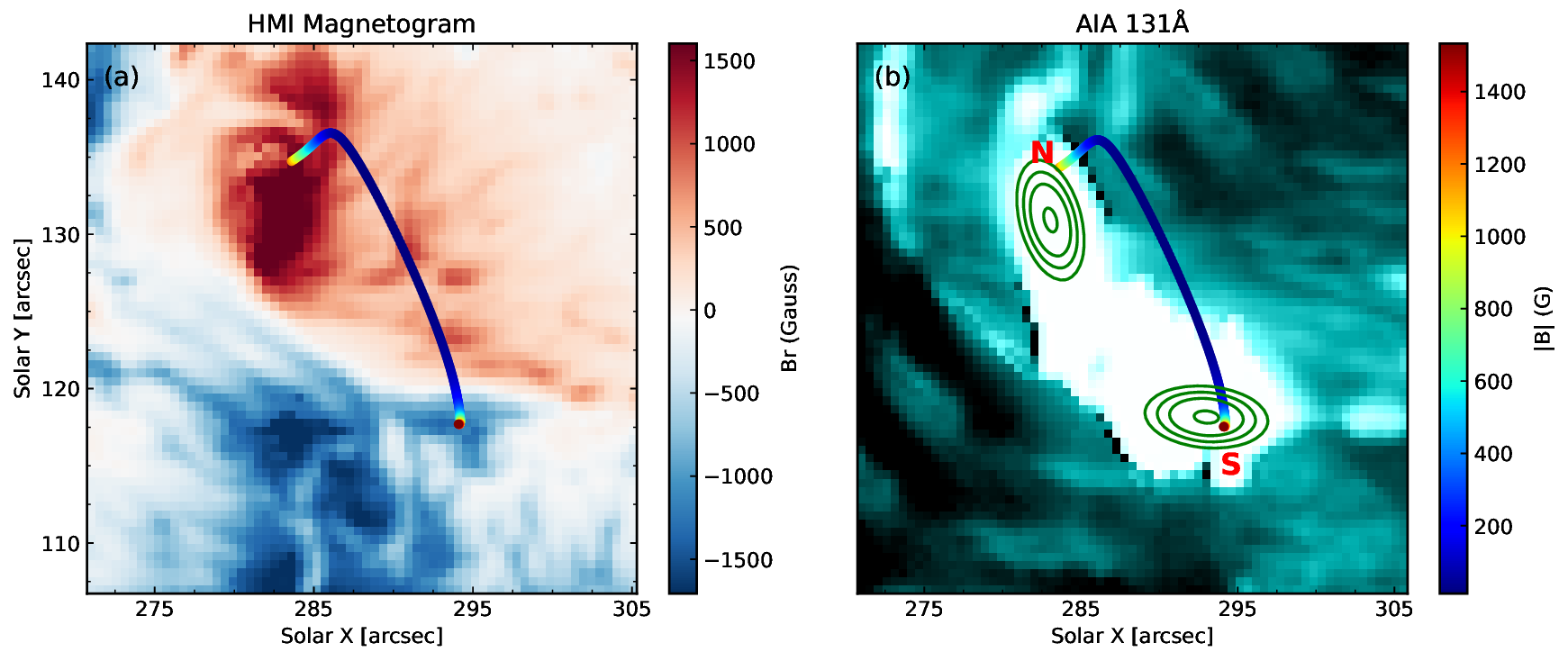}}
\caption{\textit{Magnetic configuration and electron transport geometry.}
(a) The HMI magnetogram of AR 11283 following the X2.1 flare on September 6, 2011 is shown. (b) The SDO/AIA 131~\AA~image, illustrating the correlation between the extrapolated field line and the observed ribbon morphology. The NLFFF extrapolation is over-plotted on both the SDO/HMI and SDO/AIA data, with the colour in panel (b) indicating the field line strength in Gauss. The red signs of $N$ and $S$ represent the northern and southern parts of the loop.
}
          \label{fig1}
 \end{figure*}

The active region investigated in this study is NOAA AR 11283, which produced the X2.1-class flare on 2011 September 6. This active region has been extensively studied due to its complex magnetic topology and associated flare activity (e.g., \cite{Liu2014,Yousefzadeh2021,Yousefzadeh2022}. 
The NLFFF extrapolation is constructed using the HMI vector magnetogram observed at 22:24 UT (see Figure~\ref{fig1}a), corresponding to the post-impulsive phase configuration of the flare. Here, the NLFFF equations are solved using an optimisation approach with a weighting function following the method developed by \citet{Wiegelmann2004}, providing a three-dimensional magnetic configuration suitable for particle transport simulations.
The NLFFF approximation assumes that the coronal plasma is sufficiently low-$\beta$ above the chromosphere, such that the Lorentz force is approximately balanced and the magnetic field can be considered force-free.
The extrapolated magnetic field is then used to trace the coronal loop structure along which energetic electrons are transported. 

The reconstructed magnetic field is over-plotted on the SDO/AIA 131~\AA~image observed at 22:18:26 UT to identify the flaring loop geometry and establish the connection between the extrapolated magnetic structure and the observed ribbons morphology. The selected loop is consistent with the geometry used in the observational analysis of \citet{Krucker2021}. Following their approximation, the coronal loop is represented by a semicircular structure with an approximate foot-point separation of $25''$, while the conjugate flare ribbons have characteristic lengths of about $7''$ (e.g. see Figure~\ref{fig1}b). These observational constraints are adopted to estimate the ribbon area used later for calculating the deposited energy flux from precipitating electrons. 

The extrapolated magnetic field shows a clear asymmetry between the two loop foot-points. The northern foot-point is associated with a weaker magnetic field strength, while the southern foot-point is rooted in a stronger magnetic concentration. 
Therefore, energetic electrons transported along the loop experience asymmetric precipitation toward the conjugate ribbons. 
The magnetic configuration is assumed to be stationary during particle transport calculations because the simulated electron propagation time is much shorter than the timescale associated with the global evolution of the post-flare magnetic structure.

\subsection{Guiding-center transport model}
The energetic electron transport is efficiently calculated using a three-dimensional GCA model, as the electron gyro-radius is significantly smaller than the coronal magnetic field's characteristic spatial scale.
This approach efficiently simulates the evolution and transport of millions of electrons by removing rapid gyro-motion, while retaining essential parallel and drift motions to preserve the physics governing electron propagation along realistic coronal magnetic fields. 
The guiding-center motion is governed by (e.g. \cite{Northrop1963})

\begin{equation}
\frac{\mathrm{d}\mathbf{r}}{\mathrm{d}t}
=
\mathbf{u}
+
v_{\parallel}\mathbf{b},
\end{equation}

where $\mathbf{r}$ denotes the guiding-center position, $v_{\parallel}$ is the particle velocity parallel to the magnetic field, and $\mathbf{b}=\mathbf{B}/B$ is the unit vector along the magnetic field.

The total perpendicular drift velocity is

\begin{align}
\mathbf{u}
=&\
\mathbf{u}_E
+
\frac{m}{q}
\frac{v_{\parallel}^{2}}{B}
\left[
\mathbf{b}\times
(\mathbf{b}\cdot\nabla)\mathbf{b}
\right]
+
\frac{m}{q}
\frac{\mu}{B}
\left[
\mathbf{b}\times\nabla B
\right]
\nonumber\\
&
+
\frac{m}{q}
\frac{v_{\parallel}}{B}
\left[
\mathbf{b}\times
(\mathbf{b}\cdot\nabla)\mathbf{u}_E
\right]
+
\frac{m}{q}
\frac{v_{\parallel}}{B}
\left[
\mathbf{b}\times
(\mathbf{u}_E\cdot\nabla)\mathbf{b}
\right]
\nonumber\\
&
+
\frac{m}{q}
\frac{1}{B}
\left[
\mathbf{b}\times
(\mathbf{u}_E\cdot\nabla)\mathbf{u}_E
\right],
\end{align}

where

\begin{equation}
\mathbf{u}_E
=
\frac{\mathbf{E}\times\mathbf{B}}{B^{2}}
\end{equation}

is the $\mathbf{E}\times\mathbf{B}$ drift, $m$ and $q$ denote the particle mass and charge, respectively, and $\mu$ is the first adiabatic invariant.

The evolution of the parallel velocity satisfies

\begin{align}
\frac{\mathrm{d}v_{\parallel}}{\mathrm{d}t}
=
&
\frac{q}{m}
\mathbf{E}\cdot\mathbf{b}
-
\mu
(\mathbf{b}\cdot\nabla B)
+
v_{\parallel}
\mathbf{u}_E
\cdot
(\mathbf{b}\cdot\nabla)\mathbf{b}
\nonumber\\
&
+
\mathbf{u}_E
\cdot
\left[
(\mathbf{u}_E\cdot\nabla)\mathbf{b}
\right],
\end{align}

Here, the first term describes acceleration by the electric field, the second term represents the magnetic mirror force, and the remaining terms account for inertial corrections associated with magnetic curvature and drift motion. 
The presented model simplifies analysis by omitting the electric field parallel to magnetic field lines, as its influence on particle propagation outside the acceleration region is negligible. This eliminates the need for electric field and $\mathbf{E}\times\mathbf{B}$ drift terms. This approach can efficiently describe energetic electron propagation in a prescribed static magnetic configuration, focusing on particle transport without self-consistent plasma feedback.

The injected electron population forms an anisotropic beam with average velocities of \(v_{\parallel}\simeq0.23c\) parallel and \(v_{\perp}\simeq0.15c\) perpendicular to the local magnetic field. The corresponding total speed is \(v \simeq0.28c\), consistent with an electron kinetic energy of approximately \(20~\mathrm{keV}\). For the adopted velocity components, the averaged characteristic initial pitch angle is \(\alpha_{0}\simeq35^{\circ}\) relative to the local magnetic field.
Energetic electrons are injected near the loop-top region using an observationally motivated beam distribution, representing acceleration during the impulsive phase of the flare. 
The initial beam properties, including the characteristic energy distribution, are chosen to remain consistent with the non-thermal electron population inferred from RHESSI observations of the X-class flare analysed by \citet{Krucker2021}. 
In the modelled loop, electrons with an average energy of \(20~\mathrm{keV}\) have a collision-less transit time of approximately \(0.35~\mathrm{s}\) across the characteristic field-line length.
To accurately simulate the gradual approach to a quasi-steady precipitation rate, a longer simulation time is needed to capture repeated magnetic reflections, pitch-angle diffusion, and loss-cone filling.
Here, the total numerical transport time is ($\sim$$10~\mathrm{s}$), which is significantly shorter than the global evolution time of the post-flare magnetic structure.

During propagation, turbulent pitch-angle scattering is incorporated using a stochastic integration method for the focused transport equation. Stochastic integration methods provide a particle-based representation of the corresponding Fokker--Planck description \citep{Zhang1999}, and related approaches have been applied to energetic-electron transport in solar flares \citep{Kong2022,Yousefzadeh2021}.

The scattering timescale (\(\tau_{\mathrm{sc}}=\nu_{\mathrm{sc}}^{-1}\)) represents an effective angular-decorrelation time. It characterises the interval over which the cumulative effect of many unresolved microscopic interactions produces a stochastic change in the electron pitch angle. The chosen timescale controls how often these random perturbations are applied, and the isotropic distribution is the stationary angular distribution adopted for this cumulative scattering prescription, consistent with the strong-scattering limit of pitch-angle diffusion theory \citep{Kontar2014,Bian2018}.

At each numerical time step $\Delta t$, a scattering update is selected with probability

\begin{equation}
P_{\mathrm{sc}} = 1-\exp\left(-\frac{\Delta t}{\tau_{\mathrm{sc}}}\right),
\label{eq:scattering_probability}
\end{equation}

which reduces to $P_{\mathrm{sc}}\simeq \Delta t/\tau_{\mathrm{sc}}$ for $\Delta t\ll\tau_{\mathrm{sc}}$. When an update occurs, the electron kinetic energy is conserved, while the net angular decorrelation accumulated over the effective scattering interval is represented by sampling the outgoing pitch-angle cosine $\mu'$ from the stationary isotropic distribution. The corresponding velocity components are

\begin{equation}
v_{\parallel}'=v\mu', \qquad v_{\perp}'=v\sqrt{1-\mu'^2}, \qquad v^2=v_{\parallel}'^{\,2}+v_{\perp}'^{\,2}.
\label{eq:velocity_scattering}
\end{equation}

A numerical update therefore represents the cumulative effect of many unresolved microscopic scattering interactions and is not interpreted as a single collision or a single wave--particle interaction. This can significantly change $\mu$ without implying that one physical interaction reverses the electron. The resulting particle distribution and precipitation flux are determined by the collective evolution of the particle ensemble over successive decorrelation intervals.

The values $\tau_{\mathrm{sc}}=\infty,\;2,\;1,\;0.5~\mathrm{s}$ are used to investigate the transition from scatter-free propagation to increasingly rapid turbulent angular decorrelation. The limit $\tau_{\mathrm{sc}}=\infty$ corresponds to the absence of turbulent pitch-angle scattering, while decreasing $\tau_{\mathrm{sc}}$ represents more rapid cumulative decorrelation of the particle propagation direction.

Four characteristic scattering timescales, $\tau_{\mathrm{sc}}= \infty$, $2$, $1$, and $0.5~\mathrm{s}$, are investigated (smaller $\tau_{\mathrm{sc}}$ corresponds to stronger scattering) to examine how turbulent diffusion modifies electron precipitation and the associated energy deposition at the flare ribbons; the case of $\tau_{\mathrm{sc}}= \infty$ corresponds to purely collision-less transport without turbulent scattering. 
 The corresponding scattering mean free path relative to the loop length (\(\lambda_{\mathrm{sc}}/L\)) is approximately \(5.8\), \(2.9\), and \(1.4\) for \(\tau_{\mathrm{sc}}=2\), \(1\), and \(0.5~\mathrm{s}\), respectively. These cases span weak scattering to a moderate transport regime, in which the electrons travel over a substantial fraction of the loop before their pitch angles are significantly redistributed. Thus, the parameter $\tau_{\mathrm{sc}}$ controls the frequency of stochastic pitch-angle redistribution and measures the turbulent scattering strength.
  
\begin{figure*}
    \resizebox{\hsize}{!}
   {\includegraphics {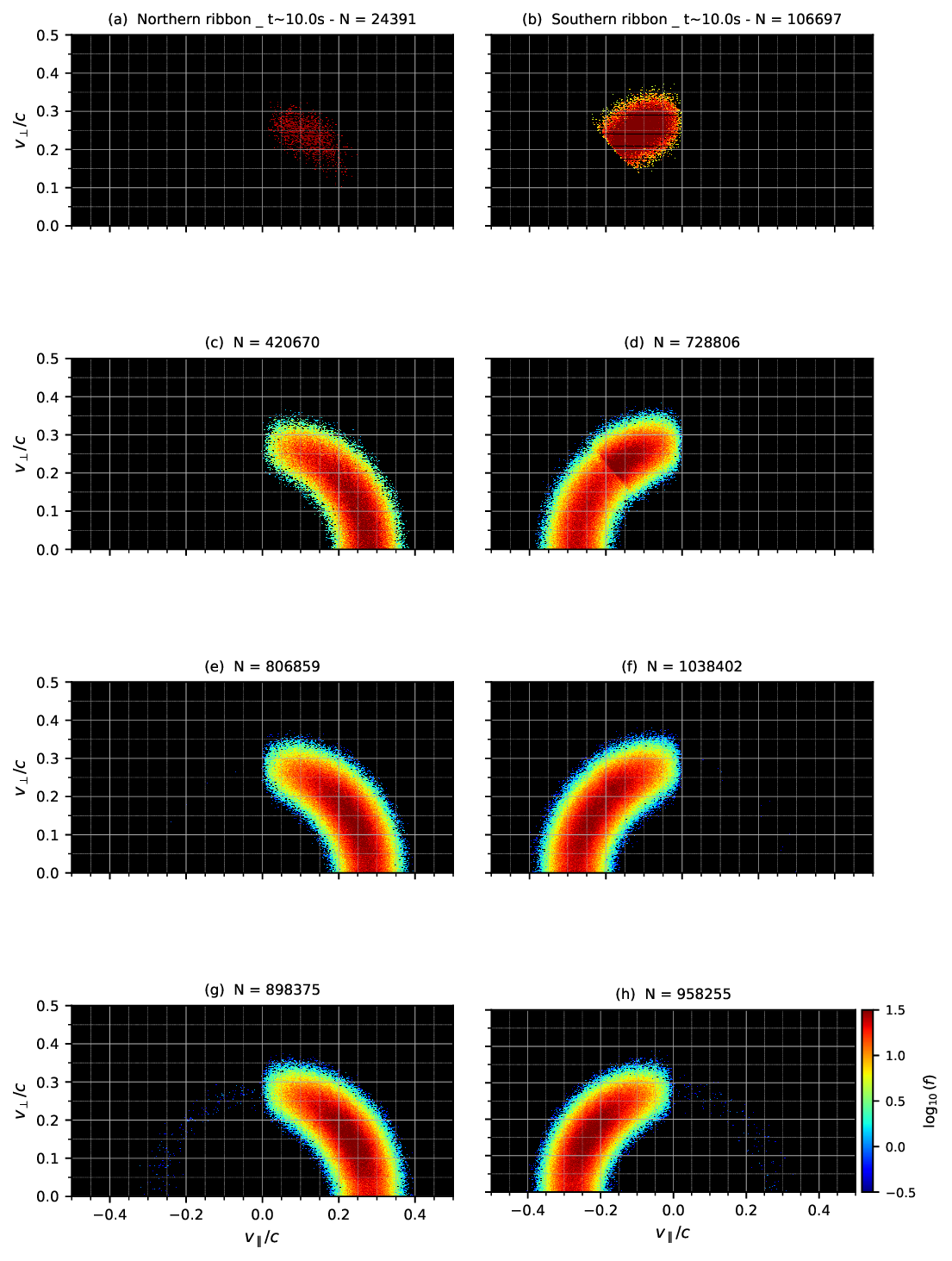}}
\caption{\textit{Electron velocity distributions at the conjugate flare ribbons.}
The velocity distribution functions of energetic electrons at the final simulation time ($\sim10~\mathrm{s}$) illustrate the time-integrated distributions of the precipitated electrons for four scattering timescales. The left column (panels (a), (c), (e), (g)) displays the North ribbon, while the right column (panels (b), (d), (f), (h)) shows the South ribbon. Each row represents a different scattering condition, with $\tau_{\mathrm{sc}}= \infty$, $2$, $1$, and $0.5~\mathrm{s}$ plotted from top to bottom.}
          \label{fig2}
 \end{figure*}

Once a particle reaches either ribbon boundary, it is considered to have precipitated into the lower atmosphere and is subsequently removed from the computational domain, so that only newly arriving particles contribute to the precipitation rate during each successive time interval. 
The resulting electron velocity distribution functions at different ribbon locations provide the kinetic input for evaluating the precipitation properties at both flare ribbons.
The final velocity distribution functions accumulated at the northern and southern ribbons after approximately $10~\mathrm{s}$ of transport are presented in Figure~\ref{fig2}. The left column (panels (a), (c), (e), and (g)) shows the northern ribbon, whereas the right column (panels (b), (d), (f), and (h)) presents the corresponding southern ribbon for the four investigated turbulent scattering timescales. These velocity distribution functions are calculated at a chromospheric level, approximately $1500~km$ above the solar surface (\cite{Forbes1989}), defining the precipitation boundary in our GCA model. These distributions illustrate the cumulative population of electrons that precipitate into each ribbon during the entire simulation and provide the basis for evaluating how turbulent scattering modifies the electron transport, precipitation efficiency, and subsequent ribbon heating discussed in the following section.

The instantaneous deposited power at each ribbon is calculated from the total kinetic energy carried by all newly precipitating electrons during a time interval $\Delta t$,

\begin{equation}
P(t) = \sum_i \frac{E_i}{\Delta t},
\end{equation}

where $E_i$ denotes the kinetic energy of the $i_{th}$ precipitating electron. The respective deposited energy flux is calculated by dividing the deposited power by the ribbon area.
These quantities provide the primary connection between the simulated electron transport and the observational estimates of non-thermal energy deposition.

The simulation produces the temporal evolution of the precipitation rate, cumulative precipitated electrons, ribbon asymmetry, deposited power, deposited energy flux, and the energy distribution $f(E_k)$ at both flare ribbons. These quantities form the basis for the quantitative comparison with the RHESSI and EUV observations presented in Section~\ref{sec:results}.

\section{Results}
\label{sec:results}
In this section, we investigate how turbulent pitch-angle scattering modifies the transport of energetic electrons and the resulting energy deposition into the flare ribbons. We first examine the evolution of electron precipitation along the reconstructed coronal loop for different scattering timescales. We then quantify the corresponding deposited power and energy flux at the conjugate foot-points and compare the simulation results with the observational constraints derived by \citet{Krucker2021}. Finally, we discuss the implications of the simulated ribbon asymmetry for interpreting the observed energy partition between the two flare ribbons.

\subsection{Electron transport and precipitation}

\begin{figure*}[ht!]
    \resizebox{\hsize}{!}
             {\includegraphics {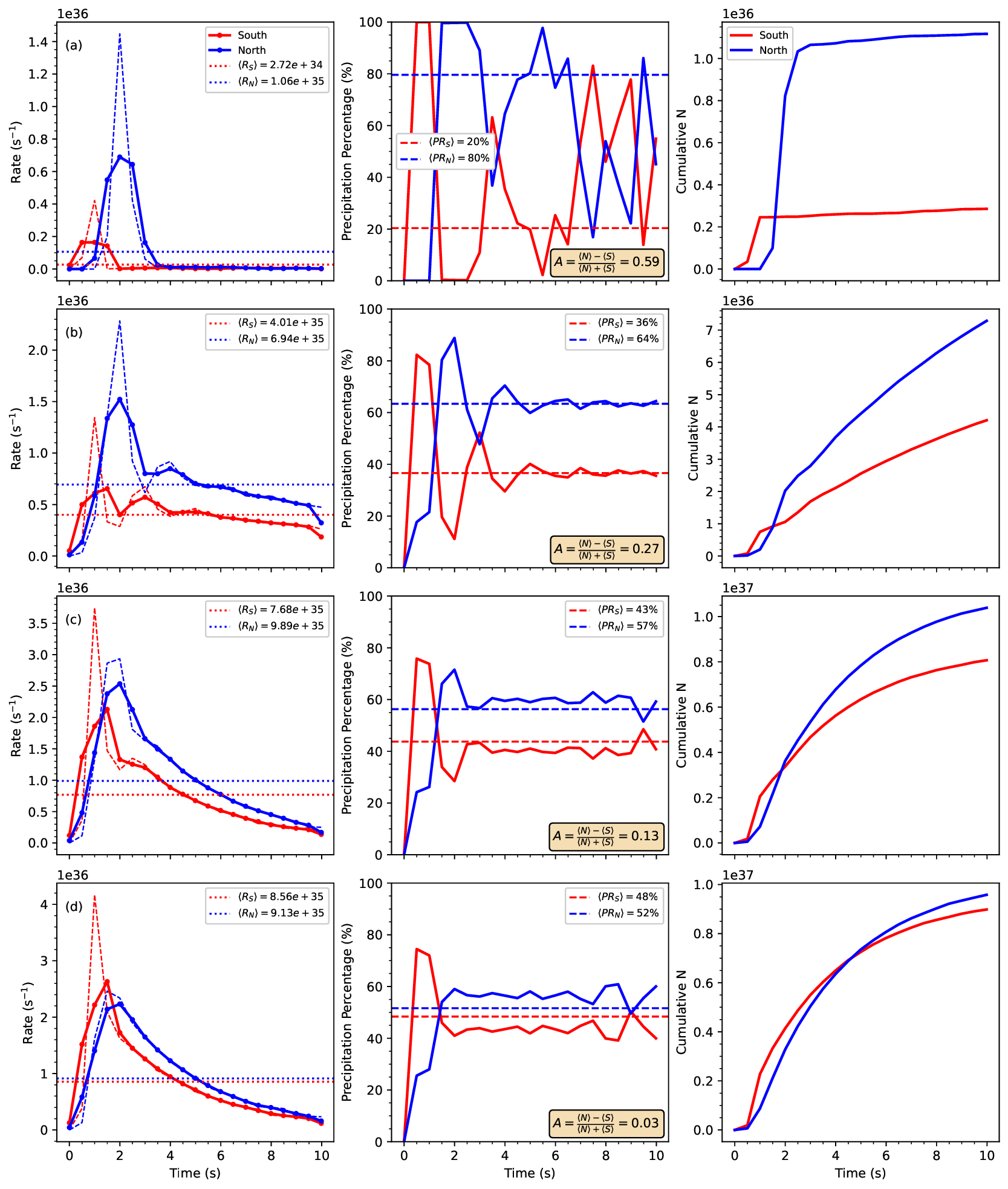}}
\caption{\textit{Precipitation and ribbon asymmetry.}
The North–South temporal evolution of precipitation rate, ribbon asymmetry, and the cumulative number of precipitating electrons are presented for four scattering timescales ($\tau_{\mathrm{sc}}= \infty, 2, 1, \text{ and } 0.5~\mathrm{s}$, panels a-d). The columns show precipitation rate, relative percentage of precipitating electrons per ribbon, and cumulative precipitation, respectively. The asymmetry parameter ($A$) is also presented for different scattering timescales.
These quantities are normalised using the observationally constrained statistical weight. The dashed line shows raw data, while the solid line shows its moving-average smoothed value in the first column.}
          \label{fig3}
 \end{figure*}

The transport of energetic electrons from the loop-top toward the conjugate flare ribbons is controlled by the combined effects of magnetic mirroring and turbulent pitch-angle scattering. Figure~\ref{fig3} presents the temporal evolution of the precipitation rate, ribbon asymmetry, and cumulative precipitation for four different scattering timescales, where each row corresponds to $\tau_{\mathrm{sc}}= \infty$, $2~\mathrm{s}$, $1~\mathrm{s}$, and $0.5~\mathrm{s}$ (panels (a)--(d), respectively). The three columns display the precipitation rate, precipitation asymmetry, and cumulative number of precipitating electrons.

To connect numerical results with the physical flare system, each simulated particle is assigned a statistical weight, $w$. This weight scales the simulated population to the actual non-thermal electron population, which is inferred from the observed energy deposition rate. Instead of arbitrary normalisation, $w$ is chosen so that the total simulated beam matches the non-thermal electron content derived from RHESSI spectroscopy.
This deposited power is computed as

\begin{equation}
P_{\mathrm{sim}}
=
\frac{w}{\Delta t}
\sum_{i=1}^{N_{\mathrm{prec}}}
E_i,
\end{equation}

$N_{\mathrm{prec}}$ is the newly precipitating electrons during the time interval $\Delta t$ with $E_i$ being the kinetic energy of the $i_{th}$ precipitating electron. By requiring the simulated deposited power to reproduce the observed non-thermal energy deposition rate reported by \citet{Krucker2021},

\begin{equation}
P_{\mathrm{sim}}
=
P_{\mathrm{obs}}
\approx
6.2\times10^{28}
~\mathrm{erg~s^{-1}},
\end{equation}

the corresponding statistical weight is estimated as

\begin{equation}
w=
\frac{P_{\mathrm{obs}}\Delta t}
{\sum_iE_i},
\end{equation}

which is nearly of the order of $10^{31}$ for the present simulations of two million electrons with an average energy of approximately 20 $keV$.
This observation-driven normalisation enables the simulated precipitation history to be directly compared with the inferred electron energy deposition in the observed flare. Consequently, all precipitation rates, deposited powers, and particle numbers reported below correspond to physically scaled quantities constrained by the observed flare energetics.

Using this observation-constrained normalisation, the average number of precipitating electrons over the $\sim10~\mathrm{s}$ simulation interval was determined for each scattering timescale. In the absence of turbulent scattering, approximately $10^{35}$ electrons precipitate into the North ribbon, while only $2.7\times10^{34}$ reach the South ribbon. Introducing turbulent scattering substantially increases the total precipitation. For $\tau_{\mathrm{sc}}=2~\mathrm{s}$, the precipitating populations increase to approximately $7\times10^{35}$ and $4\times10^{35}$ electrons for the North and South ribbons, respectively. The largest precipitation is obtained for $\tau_{\mathrm{sc}}=1~\mathrm{s}$, where approximately $10^{36}$ electrons precipitate into the North ribbon and $8\times10^{35}$ into the South ribbon. A comparable total precipitation is obtained for $\tau_{\mathrm{sc}}=0.5~\mathrm{s}$, yielding approximately $9\times10^{35}$ and $8.5\times10^{35}$ electrons in the North and South ribbons, respectively. These values are shown in the first column of Figure~\ref{fig3} for different time scale of scattering.

The absence of scattering allows electrons to maintain pitch angles, resulting in coherent transport dynamics influenced by magnetic focusing and mirroring. The weaker magnetic field at the North foot-point rooted in a larger loss cone, causing approximately $80\%$ of energetic electrons to precipitate into the North ribbon, compared to $20\%$ in the South. 
The different magnetic mirror strengths at conjugate foot-points result in asymmetric precipitation rates. As electrons experience mirroring effect and bounce repeatedly within the loop, the precipitation is not evenly distributed in time; instead, the results show peak precipitation occurring at one foot-point at a certain time, followed by the other. 
Consequently, the cumulative precipitation reveals a strong, nearly fourfold imbalance between the two foot-points without the effect of scattering (e.g., see the first row of Figure~\ref{fig3}).

The effect of increasing the turbulent scattering significantly modifies this transport. As the scattering frequency increases, pitch-angle diffusion continuously repopulates the loss cone, allowing a larger fraction of previously trapped electrons to escape toward both foot-points. Therefore, the precipitation becomes progressively more symmetric. The South ribbon receives approximately $36\%$, $43\%$, and $48\%$ of the precipitating electrons for $\tau_{\mathrm{sc}}=2~\mathrm{s}$, $1~\mathrm{s}$, and $0.5~\mathrm{s}$, respectively, while the North ribbon receives the remaining $64\%$, $57\%$, and $52\%$. 
The precipitation asymmetry between conjugate foot-points, quantified by $A = (N_{\rm N} - N_{\rm S})/(N_{\rm N} + N_{\rm S})$, decreases monotonically with increasing scattering frequency: averaged values of $A = 0.59$, $0.27$, $0.13$, and $0.03$ are obtained for $\tau_{\rm sc} = \infty$, $2$, $1$, and $0.5$~s, respectively.
The gradual transition from magnetic mirroring to turbulent diffusion shows that scattering randomises pitch angles, causing a loss of directional memory and destroying coherent bouncing patterns. This results in precipitation becoming a smoother stochastic process, which in turn smooths the asymmetry curve (e.g., Figure~\ref{fig3}, second column).

The cumulative number of precipitating electrons, represented by the third column in Figure~\ref{fig3}, is calculated by summing the number of newly precipitating electrons ($N_{\mathrm{prec}}(t_i)$) during each simulation interval. This cumulative value represents the total number of electrons deposited into each ribbon from the beginning of the simulation up to that specific time, with the final values reflecting the total over the entire ~10 s evolution. The figure specifically illustrates the total number of precipitated electrons at 0.5 s intervals until the simulation concludes.

 To establish the limiting behaviour at stronger scattering, an additional test was performed with \(\tau_{\mathrm{sc}}=0.15~\mathrm{s}\), for which \(\lambda_{\mathrm{sc}}/L \simeq 0.4\). In this regime, the scattering mean free path is shorter than the loop length and the transport approaches the strong-diffusion limit. 
The final precipitation partition becomes nearly symmetric, with approximately \(51\%\) of the precipitating electrons reaching the northern ribbon and \(49\%\) reaching the southern ribbon. In addition to this stronger symmetry, the total precipitation rate is slightly reduced compared with the \(\tau_{\mathrm{sc}}=0.5~\mathrm{s}\) case, because strong pitch-angle scattering prevents the propagation of electrons from the injection point to the foot-points. 
We do not discuss this limiting case further because the primary aim of this work is to identify the parameter range that reproduces both the observed non-thermal energy-deposition scale and the ribbon asymmetry.

These results show that turbulent scattering redistributes energetic electrons between conjugate flare ribbons, increasing the total number that escape magnetic trapping and deposit energy in the chromosphere. Strong pitch-angle scattering can largely remove the north--south asymmetry produced by unequal magnetic convergence at the two foot-points, but reduces overall precipitation efficiency. This process regulates non-thermal energy distribution within the flare loop. Thus, ribbon asymmetry contains information about the efficiency of post-reconnection pitch-angle scattering, potentially serving as an observable diagnostic of turbulence in solar flares.

\subsection{Turbulent scattering regulates ribbon heating}

\begin{figure*}
    \resizebox{\hsize}{!}
   {\includegraphics{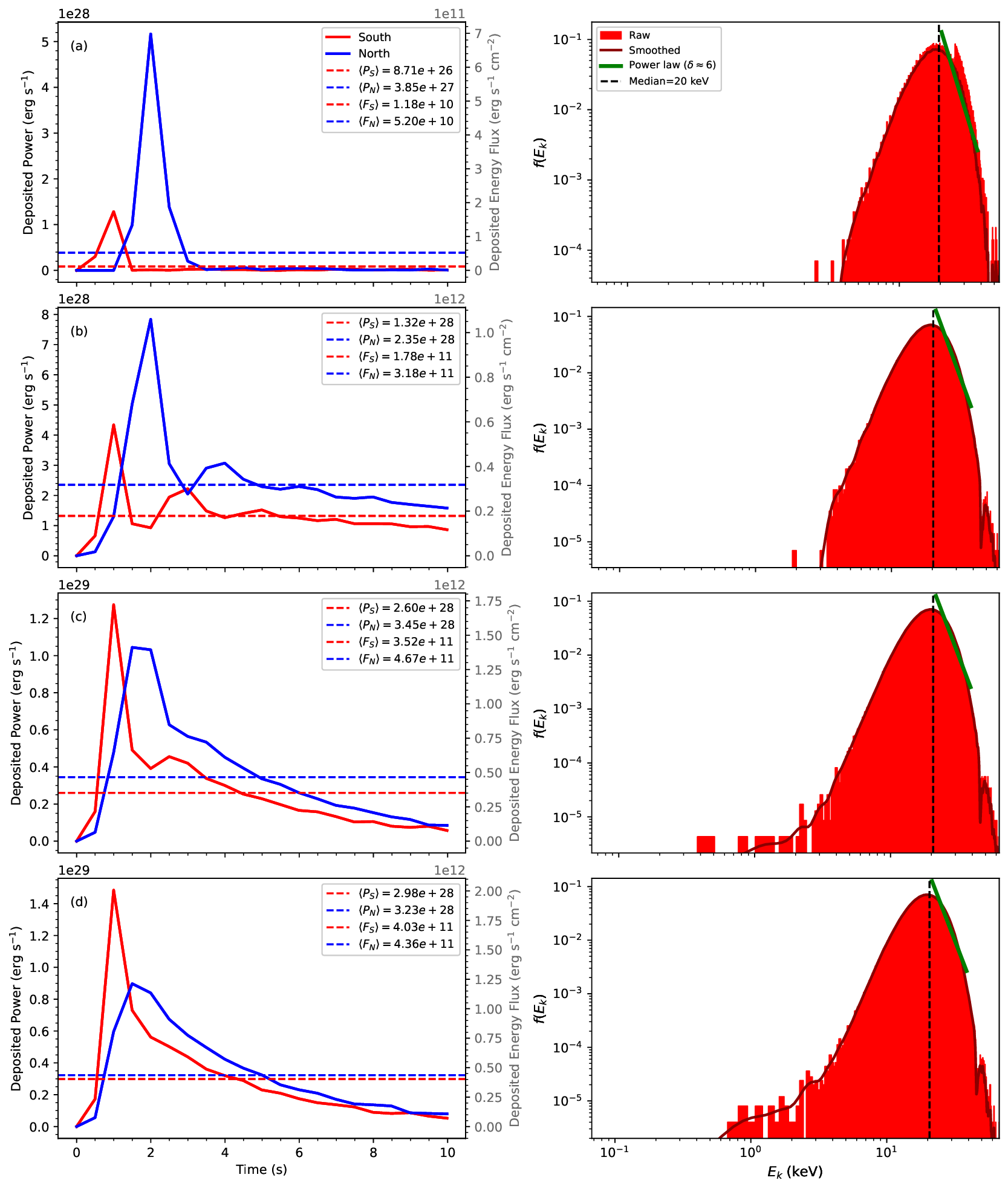}}
\caption{\textit{Energetic electron deposition and energy distributions.}
The North–South temporal evolution of the deposited power and deposited energy flux at the conjugate ribbons, together with the kinetic-energy distributions $f(E_k)$ of the precipitating populations, for the four scattering timescales ($\tau_{\mathrm{sc}}= \infty, 2, 1, \text{ and } 0.5~\mathrm{s}$, panels a-d). The first column shows deposited power and energy flux, and the second column represents the electron energy spectra. The deposited quantities are normalised using the observationally constrained statistical weight and ribbon area.}
          \label{fig4}
 \end{figure*}

The modification of electron transport directly influences the amount of non-thermal energy deposited into the chromosphere. Figure~\ref{fig4} summarises the deposited power, deposited energy flux, and electron energy spectra for the four scattering timescales. Each row again corresponds to $\tau_{\mathrm{sc}}= \infty$, $2~\mathrm{s}$, $1~\mathrm{s}$, and $0.5~\mathrm{s}$, while the columns present the deposited power, deposited energy flux, and the corresponding $f(E_k)$ distributions. The total ribbon area, as estimated from the observational paper, is approximately $7.4 \times 10^{16}~\mathrm{cm}^{2}$ (e.g. see Equation~\ref{eq:energy_flux}). This value is adopted consistently in our calculation of the deposited energy flux.
In the absence of scattering, fewer energetic electrons enter the loss cone compared to when turbulence scattering is present. The resulting total deposited power into the ribbons is approximately $4.7\times10^{27}~\mathrm{erg\,s^{-1}}$, corresponding to an energy flux of about $6.4\times10^{10}~\mathrm{erg\,s^{-1}\,cm^{-2}}$. Most electrons remain magnetically trapped within the coronal loop, limiting the transfer of non-thermal energy into the chromosphere.

Introducing turbulent scattering strongly enhances the precipitation efficiency. For $\tau_{\mathrm{sc}}=2~\mathrm{s}$, the deposited power increases to approximately $3.7\times10^{28}~\mathrm{erg\,s^{-1}}$, while the deposited energy flux reaches about $5\times10^{11}~\mathrm{erg\,s^{-1}\,cm^{-2}}$. The enhancement becomes even more pronounced for $\tau_{\mathrm{sc}}=1~\mathrm{s}$, yielding a deposited power of approximately $6\times10^{28}~\mathrm{erg\,s^{-1}}$ and an energy flux of about $8.2\times10^{11}~\mathrm{erg\,s^{-1}\,cm^{-2}}$. A comparable result is obtained for $\tau_{\mathrm{sc}}=0.5~\mathrm{s}$, where the deposited power reaches approximately $6.2\times10^{28}~\mathrm{erg\,s^{-1}}$ and the energy flux increases slightly to $8.3\times10^{11}~\mathrm{erg\,s^{-1}\,cm^{-2}}$.
Although the $1~\mathrm{s}$ and $0.5~\mathrm{s}$ cases produce comparable energy deposition, they differ noticeably in the spatial partition of the precipitating electrons. The $0.5~\mathrm{s}$ case approaches nearly symmetric precipitation between the two ribbons, whereas the $1~\mathrm{s}$ case preserves a moderate asymmetry that closely resembles the observational estimates. This indicates that relatively weak turbulent scattering provides the most realistic balance between efficient precipitation and the spatial distribution of deposited energy (e.g. see the first column of Figure~\ref{fig4}).

The corresponding electron energy spectra shown in the second column of Figure~\ref{fig4} further indicate that turbulent scattering primarily modifies the precipitation efficiency rather than substantially altering the overall energy distribution of the injected electrons.
The spectral index ($\gamma \approx 2.9$) reported by \citet{Krucker2021} describes the observed hard-X-ray photon spectrum. 
In contrast, the last column of Figure~\ref{fig4} shows a larger index ($\delta \approx 6$), which is derived from fitting the kinetic-energy distribution of simulated precipitating electrons.
 These indices are not directly comparable as they describe different distributions. While the simplest non-relativistic collisional thick-target approximation suggests, $\gamma= \delta-1$, meaning an observed photon index of $\gamma \approx 2.9$ would correspond to an injected electron index of $\delta \approx 3.9$, a direct comparison necessitates forward modelling of the hard-X-ray photon spectrum from the simulated electron distribution, accounting for target density and energy-loss physics. Thus, Figure~\ref{fig4} characterises the modelled precipitating electrons rather than directly reproducing the RHESSI photon spectral index.

\subsection{Comparison with flare ribbon observations}

To assess the realism of the model, we compare the simulated results with the observational constraints reported by \citet{Krucker2021}. Their analysis of the X-class flare inferred a non-thermal energy deposition rate of at least $\sim6.2\times10^{28}~\mathrm{erg\,s^{-1}}$ during the impulsive phase. The simulations with $\tau_{\mathrm{sc}}=1~\mathrm{s}$ and $0.5~\mathrm{s}$ produce deposited powers of approximately $6\times10^{28}$ and $6.2\times10^{28}~\mathrm{erg\,s^{-1}}$, respectively, both in agreement with the observational estimate. Likewise, the corresponding deposited energy fluxes of approximately $8.2\times10^{11}~\mathrm{erg\,s^{-1}\,cm^{-2}}$ are consistent with the energy required to produce intense flare ribbon heating.

An additional interesting result emerges from the ribbon asymmetry. \citet{Krucker2021} reported hot ribbon emission measures of approximately $2.5\times10^{48}~\mathrm{cm^{-3}}$ for the North ribbon and $1.9\times10^{48}~\mathrm{cm^{-3}}$ for the South ribbon. These values correspond to relative contributions of approximately $57\%$ and $43\%$, respectively. The simulation with $\tau_{\mathrm{sc}}=1~\mathrm{s}$ predicts nearly the same precipitation partition, with approximately $57\%$ of the precipitating electrons reaching the North ribbon and $43\%$ reaching the South ribbon (Figure~\ref{fig3}c, middle panel). 
The close correspondence between the modelled partition of precipitating electrons and the observed partition of hot-ribbon emission measure suggests that the transport model properly represent the spatial distribution of flare energy deposition between conjugate flare ribbons, even though the emission measure reflects the plasma's thermal response rather than direct electron precipitation. 
This agreement should not be interpreted as a one-to-one correspondence between precipitating electrons and emission measure, because the latter depends on the subsequent heating, evaporation, conduction, and radiative evolution of the ambient plasma. Nevertheless, it motivates a more complete treatment in which the time-dependent energy flux obtained from the guiding-center simulations is used as the input to a model of the thermal response of the lower atmosphere.

\section{Discussion}
\label{sec:Discussion}

The present model is designed to isolate the role of magnetic geometry and turbulent pitch-angle scattering in controlling energetic-electron precipitation. Coulomb scattering and collisional energy losses are therefore not included in the guiding-center transport calculation. This approximation is valid when the coronal density is low enough for the Coulomb timescale to exceed the characteristic turbulent scattering timescale. 
Coulomb scattering times for $20~\mathrm{keV}$ electrons are estimated around $8~\mathrm{s}$ for an electron density of $n_e=10^{9}~\mathrm{cm^{-3}}$, increasing as density decreases (\cite{Aschwanden2002}). This duration is significantly longer than the turbulent scattering times explored here ($\tau_{\mathrm{sc}}=0.5$--$2~\mathrm{s}$), suggesting that turbulent scattering can dominate the pitch-angle evolution when the coronal density is below approximately $10^{9}~\mathrm{cm^{-3}}$. This is consistent with general AIA differential emission measure studies of coronal loops \citep{Aschwanden2013, Guennou2012}). 
However, the coronal density during the impulsive phase is not sufficiently well constrained to establish this condition directly. \citet{Krucker2021} inferred densities of order $10^{11}~\mathrm{cm^{-3}}$ for the hot flare-loop plasma, but this estimate characterises the strongly heated plasma during the impulsive phase and does not directly constrain the lower-density post-flare coronal configuration represented in the present calculation. Independent estimates of the coronal density would therefore be valuable for further assessing the relative importance of Coulomb and turbulent scattering.

\cite{Kuznetsov2026} have performed Fokker-Planck simulations of electron propagation in a well-observed solar flare, and computed the corresponding microwave emission; the Coulomb scattering and scattering on turbulence have been considered. They have found that without turbulence, the electrons experience a long trapping, which is inconsistent with the available microwave observations. In their study, the best agreement with the observations has been achieved for the turbulence parameters providing a moderate diffusion regime, when the mean free path of the electrons with the energies of about one MeV is comparable to the length of the flaring loop.

We track electrons through the corona, terminating calculations above the chromosphere where precipitation is considered a boundary loss. Thus, our transport calculations focus on the coronal loop, and the resulting quantity represents the kinetic power delivered to the precipitation boundary rather than a complete calculation of chromospheric heating. In particular, the present model does not include the return-current electric field that can develop in response to the net field-aligned current carried by energetic electrons. Such a field can decelerate precipitating electrons while providing an additional heating channel through Joule dissipation \citep{Varady2014, Karlicky1992}. This effect is important for high-energy electron transport into the cooler, denser lower chromosphere. In the present model, however, the precipitation boundary represents the hotter upper-chromospheric region associated with the observed $131~\AA$ flare ribbons, above these deeper layers. This boundary isolates the transport's contribution to the energy delivered to the ribbons. The possible influence of this feedback is assessed below through a post-processing estimate of the net field-aligned current.

To assess the possible influence of the return current on energetic-electron transport, we performed a post-processing estimate of the net beam current density at four sampling locations distributed along one leg of the coronal loop, from the loop top to the vicinity of the foot-point. We excluded the initial injection phase and calculated the net current by summing the signed parallel velocities of all energetic electrons present in each sampling region. The initial injection phase was excluded from our analysis because it primarily serves as a numerical preparation for the subsequent evolution, which is the focus of our study. 
The effective sampling cross-section was estimated using magnetic-flux conservation, $B(s)A(s)\simeq\mathrm{constant}$. Assuming approximately equal areas for the two observed flare ribbons, the area associated with one foot-point is estimated to be $A_{\rm fp}\simeq3.7\times10^{16}\ {\rm cm^2}$, based on the total observed ribbon area. Using the magnetic-field strengths at the foot-point ($B_{\rm fp}\simeq1550$ G) and loop top ($B_{\rm LT}\simeq50$ G), the corresponding loop-top area is estimated to be $A_{\rm LT}\simeq1.15\times10^{18}\ {\rm cm^2}$. We approximate the average cross-sectional area along one loop leg by averaging these endpoint values, giving ($A_{\rm mean}\simeq6\times10^{17}\ {\rm cm^2}$). This provides an approximate effective area for the current-density estimates rather than a spatially resolved reconstruction of the flux-tube geometry.

The calculated net current includes both directions of energetic-electron propagation and therefore accounts for the partial compensation between counter-streaming populations. In an ideal situation, a symmetrical magnetic loop with identical magnetic mirrors would create nearly equal currents flowing in opposite directions. However, our loop is not symmetrical and it has different magnetic field strengths at its two ends, which leads to different mirror conditions. This asymmetry, combined with pitch-angle scattering and electron loss into precipitation boundaries, leads to an imbalance in particle precipitation as the two counter-streaming populations do not completely cancel each other out.
The time-averaged absolute net current densities depend on both the scattering timescale and the sampling location. Across the four scattering cases, they range from $10^7$ to $1.3\times10^9$ ${\rm statA\,cm^{-2}}$ near the loop top, from $1.4\times10^8$ to $9\times10^8\,{\rm statA\,cm^{-2}}$ in the lower area, from $2\times10^8$ to $2.2\times10^8\,{\rm statA\,cm^{-2}}$ below that, and from $4.2\times10^6$ to $2.4\times10^8\,{\rm statA\,cm^{-2}}$ near the foot-point. The degree of compensation also varies spatially: it is generally high near the loop top and intermediate locations, whereas a larger imbalance between oppositely directed populations occurs near the foot-point in the scattering cases. The stronger residual current near the foot-point in the scattering cases is consistent with the progressive loss of one component of the counter-streaming population into the precipitation boundary, while the unequal magnetic-field strengths at the two foot-points produce an overall asymmetric precipitation pattern.

We further estimated the corresponding return-current drift velocity and classical Spitzer electric field, adopting representative hot flare-coronal conditions of $T_e=20$ MK, $n_e=5\times10^{10}\ {\rm cm^{-3}}$, and a Coulomb logarithm of 20. The adopted temperature is intended as a representative condition for the hot post-flare corona rather than as a spatially resolved measurement. For the four sampling locations, the estimated drift velocities range from $4.4\times10^5$ to $5.8\times10^7\ {\rm cm\,s^{-1}}$ near the loop top, from $5.9\times10^6$ to $3.7\times10^7\ {\rm cm\,s^{-1}}$ in the lower area, from $8.6\times10^6$ to $9\times10^6\ {\rm cm\,s^{-1}}$ below that, and from $1.8\times10^5$ to $10^7\ {\rm cm\,s^{-1}}$ near the foot-point. 
The corresponding classical electric-field magnitudes range approximately from $2.3\times10^{-11}$ to $3.1\times10^{-9}\ {\rm statV\,cm^{-1}}$, $3.1\times10^{-10}$ to $2\times10^{-9}\ {\rm statV\,cm^{-1}}$, $4.6\times10^{-10}$ to $4.9\times10^{-10}\ {\rm statV\,cm^{-1}}$, and $9.3\times10^{-12}$ to $5.5\times10^{-10}\ {\rm statV\,cm^{-1}}$ at the four respective locations. These values are order-of-magnitude estimates based on a representative density rather than a spatially resolved coronal density profile. The adopted temperature and density are reference conditions for this assessment, rather than independently measured values at each sampling location.

The estimated classical electric-field magnitudes correspond to approximately \(3\times10^{-9}\)--\(10^{-6}\ {\rm V\,cm^{-1}}\). Over a characteristic loop propagation length of $\sim$$10^9$ cm, the corresponding energy loss of a \(20\) keV electron is approximately \(0.003\)--\(1\) keV, which represents at most a few percent of the characteristic energy of the precipitating electrons. This indicates that counter-streaming and scattering reduce the net beam return-current magnitude to a level at which deceleration by the classical return-current electric field is negligible. Therefore, the return-current electric field inferred from the post-processed net currents should have an insignificant impact on electron energy over the considered coronal propagation length.

The return-current drift is generally below the ion sound speed, which is approximately $c_s=\sqrt{k_{\rm B}T_e/m_p}\simeq4.1\times10^7\ {\rm cm,s^{-1}}$. This holds true for the near foot-point, lower-loop, and upper-intermediate regions, while only at the loop top does the estimated drift exceed this speed. These conclusions remain qualitatively consistent even with electron temperatures ($T_e$) ranging from 10 to 20 MK. 
Together with the decreasing estimated electric-field magnitude toward the foot-point, this suggests that the local return-current response may be less pronounced near the precipitation boundary than in the upper part of the loop under the adopted conditions. This is relevant to the present study, which primarily investigates energetic-electron precipitation and its observational consequences at the flare ribbons. The initial precipitation-rate peaks occur before the counter-streaming population is fully established and may be influenced by the absence of return-current feedback. Nevertheless, these estimates do not establish that return-current effects or associated plasma instabilities are absent, nor do they rule out local beam--plasma interactions that could modify the electron energy distribution in the corona.

The dependence on scattering is also spatially non-monotonic. Stronger pitch-angle scattering substantially reduces the net beam current in the upper loop by enhancing the compensation between oppositely directed populations. Near the foot-point, however, the residual current can increase for stronger scattering because precipitation removes particles from the two counter-streaming populations at different rates. Thus, scattering generally reduces the directional beam current in the upper loop but does not guarantee a smaller residual current at every position.

We emphasise that this post-processing estimate is not a self-consistent treatment of the return current. The present guiding-centre model follows energetic electrons in a prescribed magnetic-field configuration, including magnetic mirroring and pitch-angle scattering, but does not evolve the background electron and ion populations, current closure, or the parallel electric field. The model therefore neglects the return-current electric field, as well as any associated return-current-driven turbulence or anomalous resistivity. The electric field associated with the return current is fundamentally coupled to the particle distribution: it both reacts to and reshapes the beam current. Calculating the field from a post-processed current while keeping the particle trajectories fixed, ignores non-linear feedback. Thus, the resulting estimate would not be more self-consistent than the initial field-free calculation. A self-consistent treatment of energetic-electron transport and the background plasma response is therefore beyond the scope of the present work. The reported deposited power is therefore a baseline estimate of the kinetic energy flux to the lower atmosphere, excluding return-current deceleration or associated Joule heating, rather than the total chromospheric heating rate.

Our results therefore suggest that turbulent scattering provides an effective mechanism for regulating both the efficiency and spatial distribution of energetic-electron precipitation. Magnetic mirroring establishes the initial asymmetry imposed by the coronal magnetic geometry, whereas pitch-angle scattering modifies the fraction of particles entering the loss cones and consequently redistributes the deposited energy between the two ribbons. The existence of a scattering regime that simultaneously produces substantial precipitation, a nearly balanced ribbon partition, and an energy deposition rate comparable to the observational constraint suggests that the observed ribbon properties can provide indirect information about coronal electron transport. In this sense, the present study quantitatively links the macroscopic magnetic-field structure with microscopic transport properties of energetic electrons in the corona and the spatially resolved energy input from flare-ribbon observations. Our results further show that realistic non-thermal energy deposition rates can be obtained from particle transport alone, without introducing a self-consistent electric field or a detailed hydrodynamic response of the lower atmosphere.

The consistency between the modelled precipitation rate and the observed energy partition between the two flare ribbons supports the physical plausibility of the electron-transport model under the observational constraints adopted for this event. 
The quantitative interpreting of the thermal EUV emission requires considering the response of the ambient plasma to deposited energy input.
The next stage of this work will therefore use the time-dependent energy deposition obtained from the guiding-center simulations as a heating input to a simplified thermal model of the flare ribbons. 
 By combining the simulated heating rate with observationally constrained plasma densities  and source-volume estimates, the evolution of the plasma temperature and emission measure can be investigated and subsequently connected to differential emission measure and synthetic EUV emission. This extension will provide a direct test of whether the electron precipitation predicted by the transport model can account for the observed thermal enhancement of the flare ribbons. 
 The present results should therefore be interpreted as constraints on the kinetic energy delivered to the upper-chromospheric precipitation boundary, rather than as a direct forward model of either EUV or hard X-ray emission.

\section{Summary and conclusions}
\label{sec:conclusion}

In this study, we developed a guiding-center particle transport model to investigate the propagation and precipitation of energetic electrons from the loop-top toward the two conjugate flare ribbons. The coronal magnetic configuration was reconstructed using a non-linear force-free field extrapolation based on SDO/HMI observations, while the energetic electrons were injected near the loop-top using an observationally motivated non-thermal distribution. Their trajectories, energies, pitch angles, and positions were followed throughout the coronal loop. The spatial variation of the magnetic field naturally produces magnetic mirroring and different loss-cone conditions at the two foot-points, while turbulent transport was represented by pitch-angle scattering with a prescribed characteristic timescale.

We investigated four scattering conditions, $\tau_{\mathrm{sc}}= \infty$, $2$, $1$, and $0.5~\mathrm{s}$, and quantified the corresponding precipitation rate, cumulative number of precipitating electrons, partition of precipitation between the two ribbons, deposited power, and deposited energy flux. The numerical particle population was converted into physical quantities using a statistical weight constrained by the observationally inferred non-thermal energy input, allowing a direct comparison with the flare energetics reported by \cite{Krucker2021}. The results show that the precipitation process is strongly modified by turbulent scattering. In the absence of scattering, magnetic mirroring produces a pronounced asymmetry, with approximately $80\%$ of the precipitating electrons reaching the North ribbon and $20\%$ reaching the South ribbon. As the scattering frequency increases, this asymmetry is progressively reduced, with precipitation fractions of approximately $64\%/36\%$, $57\%/43\%$, and $52\%/48\%$ for the North/South ribbons at $\tau_{\mathrm{sc}}=2$, $1$, and $0.5~\mathrm{s}$, respectively. 
This trend reflects the progressive loss of initial directional memory and magnetic-field preference as pitch-angle scattering becomes more frequent, causing electrons to repeatedly sample both loss cones and thereby reducing the north--south imbalance. Therefore, continuous pitch-angle scattering repopulates the loss cones at both foot-points, progressively weakening the influence of asymmetric magnetic mirror ratios and leading to a more symmetric precipitation distribution.

The corresponding energy deposition exhibits a strong dependence on the scattering timescale. Without scattering, the deposited power is only approximately $4.7\times10^{27}~\mathrm{erg\,s^{-1}}$, whereas it increases to $3.7\times10^{28}$, $6\times10^{28}$, and $6.2\times10^{28}~\mathrm{erg\,s^{-1}}$ for $\tau_{\mathrm{sc}}=2$, $1$, and $0.5~\mathrm{s}$, respectively. The deposited energy flux shows a similar trend, peaking at $8.3\times10^{11}~\mathrm{erg\,s^{-1}\,cm^{-2}}$.
These results show that turbulent scattering strongly enhances the efficiency with which energetic electrons escape magnetic trapping and deliver energy to the lower atmosphere. However, the increase is not strictly monotonic with decreasing scattering timescale: the $\tau_{\mathrm{sc}}=1~\mathrm{s}$ and $0.5~\mathrm{s}$ cases produce comparable total energy deposition, while the former provides a more favourable overall agreement with the observed ribbon asymmetry.

A key result emerges from comparing the simulation with the observational constraints detailed by \cite{Krucker2021}. Their analysis inferred a non-thermal electron energy deposition rate of at least $\sim6.2\times10^{28}~\mathrm{erg\,s^{-1}}$ during the impulsive phase. The simulations with $\tau_{\mathrm{sc}}=1~\mathrm{s}$ and $0.5~\mathrm{s}$ reproduce this energy deposition rate to within the uncertainty associated with the simplified transport model. The $\tau_{\mathrm{sc}}=1~\mathrm{s}$ case is notable because it also predicts a precipitation partition of approximately $57\%$ in the North ribbon and $43\%$ in the South ribbon. The observed hot-ribbon emission measures reported by \cite{Krucker2021} are approximately $2.5\times10^{48}~\mathrm{cm^{-3}}$ and $1.9\times10^{48}~\mathrm{cm^{-3}}$ for the North and South ribbons, respectively, corresponding to nearly the same relative partition of $57\%$ and $43\%$. Although emission measure reflects the thermal response of the plasma and cannot be identified directly with the precipitating electron population, this agreement provides encouraging evidence that the modelled transport captures an important part of the observed energy partition between the conjugate flare ribbons.

Future observations closer to the Sun may provide important constraints on the physical processes represented here. The Solar Close Observations and Proximity Experiments (SCOPE) is designed to perform in-situ measurements of energetic particles and electromagnetic fields, together with remote-sensing observations of coronal magnetic and plasma structures \citep{Lin2025}. These measurements could constrain magnetic configurations, energetic-electron populations, and near-Sun wave activity, which govern particle transport and pitch-angle scattering. In particular, improved observational constraints on the near-Sun magnetic field and energetic-particle distributions could link scattering parameters to underlying turbulence and plasma conditions.

\begin{acknowledgements}
The research results was sponsored by the National Key R$\&$D Program of China No.
2022YFF0503800, NSFC grants 12633011 and 12627800. This work was supported by the Hundred Talents Program (A) of the Chinese Academy of Sciences. M.Y. acknowledges support from Youth Speciality Category grant No. ZR2024QD206. A.K. acknowledges the support by the Ministry of Science and Higher Education of the Russian Federation, and Chinese Academy of Sciences President’s International Fellowship Initiative, PIFI Group grant No. 2025PG0008. M.K. acknowledges institutional support from project RVO-67985815 and GACR grant 26-20555L.
\end{acknowledgements}

\bibliographystyle{aa} 
\bibliography{aa_manuscript} 

\FloatBarrier 
\twocolumn

\end{document}